Segregation assisted grain boundary precipitation in a model Al-Zn-Mg-Cu alloy

Huan Zhao[1], Frédéric De Geuser[2], Alisson Kwiatkowski da Silva[1], Agnieszka Szczepaniak[1], Baptiste Gault[1*], Dirk Ponge[1*], Dierk Raabe[1]

[1] Max-Planck-Institut für Eisenforschung, Max-Planck-Str. 1, 40237 Düsseldorf, Germany

[2] Univ. Grenoble Alpes, CNRS, Grenoble INP, SIMaP, F-38000 Grenoble, France

**Abstract:**

Understanding the composition evolution of grain boundaries and grain boundary precipitation at near-atomic scale in aluminum alloys is crucial to tailor mechanical properties and to increase resistance to corrosion and stress corrosion cracking. Here, we elucidate the sequence of precipitation on grain boundaries in comparison to the bulk in a model Al-Zn-Mg-Cu alloy. We investigate the material from the solution heat treated state (475°C), through the very early stages of aging to the peak aged state at 120°C and further into the overaged regime at 180°C. The process starts with solute enrichment on grain boundaries due to equilibrium segregation accompanied by solute depletion in their vicinity, the formation of Guinier–Preston (GP) zones in the solute-enriched grain boundary regions, and GP zones growth and transformation. The equilibrium segregation of solutes to grain boundaries during aging accelerates this sequence compared to the bulk. Analysis of the ~10 nm wide precipitate-free zones (PFZs) adjacent to the solute-enriched grain boundaries shows that the depletion zones are determined by (i) interface equilibrium segregation; (ii) formation and coarsening of the grain boundary precipitates and (iii) the diffusion range of solutes in the matrix. In addition, we quantify the difference in kinetics between grain boundary and bulk precipitation. The precipitation kinetics, as observed in terms of volume fraction, average radius, and number density, is almost identical next to the depletion zone in the bulk and far inside the bulk grain remote from any grain boundary influence. This observation shows that the region influenced by the grain boundaries does not extend beyond the PFZs.

# 1  Introduction

It is well understood that grain boundaries differ in structure and properties from the grain interiors. These differences cause chemical partitioning phenomena between the two regions. More specific solutes segregate to grain boundaries, driven by the Gibbs adsorption isotherm, an effect which modifies the grain boundary composition compared to the bulk. Such equilibrium segregation phenomena play an important role for transport and phase transformations [1-5], as well as for the material's strength and strain hardening [6], mechanical creep [7], liquid metal embrittlement [8, 9] and corrosion behavior [10]. Thorough investigations of solutes behavior at grain boundaries at the near-atomic scale have been reported e.g. for Fe alloys [11-15]. Detailed segregation studies were published on ultrafine grained Al alloys [16-20], better general understanding of grain boundary segregation in conventional coarse-grained Al alloys is required, particularly in high-strength Al-alloys with multiple substitutional elements. Here we address this topic for a model Al-Zn-Mg-Cu alloy, representing the essential engineering class of 7xxx ultrahigh strength Al alloys. In these materials grain boundary segregation plays an essential role for several reasons: First, they are well known to become highly susceptible to stress corrosion cracking and subject to brittle intergranular fracture under certain conditions of heat treatment [21-24]. The embrittlement has been attributed to the formation of the coarse, incoherent precipitates along grain boundaries, which leads to softening and substantially reduces the fracture resistance of these materials [25-27]. Second, interfacial segregation in these alloys is generally associated with the formation of ~100 nm wide precipitation free zones (PFZs) as the grain boundaries trap the solutes located adjacent to them [28, 29]. This well-known phenomenon is characterized by strong elemental partitioning among the solute-enriched grain boundaries, the depleted regions adjacent to them and the unaffected grain interiors [30]. Similar aspects apply to the quenched-in vacancies to which the interfaces and precipitates can act as sinks [31, 32]. This spatially highly confined compositional variation at and around grain boundaries, which is essentially driven by the Gibbs adsorption isotherm, creates very high local electrochemical and mechanical contrast, often reducing the alloy's corrosion resistance and mechanical properties [33-35]. Both phenomena are essential as further tuning of high strength Al alloys as key materials for light weight transportation solutions is limited more by the weakness of the internal interfaces rather than by insufficient strength of the grain interiors.

The precipitation sequence in the bulk of Al-Zn-Mg-Cu alloys has been studied by several groups and is well understood [36-43], in terms of the individual processes of solid solution → GP zones→ metastable η'→ stable η (MgZn$_2$). In addition, the final structure of grain boundary precipitates has been studied [44-46]. Most of the grain boundary dependent phenomena discussed in these works refer to changes in kinetics, i.e. the precipitation and coarsening phenomena observed are generally faster on the interfaces than in the bulk grains. Yet, it must also be considered that equilibrium grain boundary segregation can create locally a different composition, characterized by much higher and/or different solute content than in the grain interior. Segregation changes the local thermodynamic state at grain boundaries in a sense that it can in principle influence the sequence, composition and type of the precipitates that are formed. However, details regarding segregation, the evolution of grain boundary chemistry, the early stages of grain boundary precipitation and the formation of the PFZs in Al-Zn-Mg-Cu alloys are lacking. These knowledge gaps hinder precise understanding of the interplay between decoration, formation, growth, and kinetics of grain boundary precipitation, and hence prevent tailoring grain boundaries and avoiding the adjacent precipitation free regions for the design of materials with better mechanical and corrosion properties.

Atom probe tomography (APT) provides three-dimensional (3D), near-atomic scale analytical mapping of materials, and is uniquely positioned to measure the local chemical composition and 3D morphology of individual grain boundary and precipitates [47-51]. More intense deployment of its capabilities to investigate grain boundaries in Al-based alloys has been hindered by the lack of reliable methods to prepare site-specific specimens, i.e. APT tips containing specific features such as grain boundaries. Most APT specimens in this field have so far been prepared with the aid of a focused-ion-beam (FIB) that makes use of a Ga source. Ga implantation has however been known to cause sample embrittlement issues in Al-alloys [52, 53], as it gets readily trapped at interfaces, dislocations and grain boundaries. Also it can cause embrittlement, which is attributed to the formation of a low-temperature Al-Ga eutectic [54]. The recent development of an alternative ion source for FIBs, based on a Xe plasma, has thus been used in the current study to alleviate this stringent limitation [55].

The objective of the present work is to investigate grain boundary precipitation in comparison to bulk precipitation, from the solution heat treated state (475°C), through the very early stage of aging (120°C/0.5 hours), peak aged state (120°C/24 hours) and further into the overaged regime (120°C/24 hours +180°C/6 hours). In that context we also followed the kinetics of partitioning of the solutes from the matrix to the interfaces, the formation and growth of the precipitates in the bulk and at the grain boundaries, solute segregation to grain boundaries and the evolution of PFZs, so as to unveil the effects of segregation on grain boundary precipitation in a model Al-Zn-Mg-Cu alloy in a more holistic fashion. We made use of a parameter-free methodology to efficiently extract information from APT data such as matrix and precipitate composition, volume fraction, and number density introduced in ref. [56]. APT is combined with transmission electron microscopy (TEM), and electron backscattered diffraction (EBSD) characterization and hardness measurements to unveil the effects of segregation on grain boundary precipitation in a model Al-Zn-Mg-Cu alloy.

## 2  Materials and methods

The alloy used in this investigation is a model Al-Zn-Mg-Cu alloy. It was laboratory-cast in a vacuum induction furnace to a rectangular ingot with dimensions of 200mm×190mm×40mm. The ingot was homogenized in an Ar-atmosphere furnace for 1.5 hours at 475 °C and subsequently water quenched. Table 1 summarizes the chemical composition of the homogenized ingot measured by wet chemical analysis. Zr was added to the alloy for forming dispersoids which help refining the grain size.

*Table 1. Bulk chemical composition of a model Al-Zn-Mg-Cu alloy obtained from wet chemical analysis.*

| Alloy  | Zn   | Mg   | Cu   | Zr    | Fe    | Si    | Al   |
|--------|------|------|------|-------|-------|-------|------|
| (wt %) | 6.22 | 2.46 | 2.13 | 0.155 | 0.021 | <0.01 | Rest |

The homogenized ingot was then hot-rolled from 40 to a 3 mm sheet at 450 °C and subsequently solution heat treated for 1 hour at 475 °C and then water quenched. The quenched material was immediately aged at 120ºC for different times and subsequently water quenched to room temperature. The aging times were 0.5 hours, 2 hours and 24 hours. Aging at 120°C for 24 hours produces the peak aged state.  The peak aged sample was further aged for 6 hours at a higher temperature of 180°C to obtain overaged material. The different aging times were selected in order to follow the kinetics of precipitation according to the hardness curve [56].

APT specimens were prepared to investigate the precipitation at the grain boundaries. EBSD was first used to locate high-angle grain boundaries. APT specimens were then prepared from such regions by site-specific preparation as outlined in ref. [57] on a FEI Helios Plasma focused ion beam (PFIB) using of a Xe source on commercial silicon micro-tip coupons (Cameca Inc.) as support. Specimens were sharpened by using a 30 kV $Xe^+$ ion beam, followed by a 5 kV final cleaning step to remove regions with higher concentrations of implanted high-energy $Xe^+$ ions. APT analyses were performed on a Cameca Instrument Inc. Local Electrode Atom Probe (LEAP) 5000XS (straight flight path). The detection efficiency of this instrument is approx. 80% owing to the improved multi-channel plate detector equipped with a fine hole-array. The measurements were performed at a base temperature of 50K in voltage-pulsed mode. The pulse fraction, pulse rate and detection rate of the measurement were 20%, 250 kHz and 1% repetition rate, respectively. Datasets containing 70-150 million ions were acquired for each aging state. APT reconstruction and analysis were carried out using the Integrated Visualization and Analysis software (IVAS). The reconstruction parameters were calibrated according to the partial structural information within the dataset following the protocol described by Gault et al. [58]. TEM observations were performed on a JEOL JEM-2200 FS, operated at 200kV conducted on a PFIB-prepared sample.

## 3  Results

A typical APT dataset of the Al-Zn-Mg-Cu alloy containing a grain boundary in the as-quenched state is shown in Fig. 1. The EBSD map shows the high angle grain boundary where several APT samples were prepared from the as-quenched state. Fig. 1(b) displays a 2D desorption map directly extracted from the APT dataset, (111) poles of the top and bottom grains are indexed as the dashed circle shows.

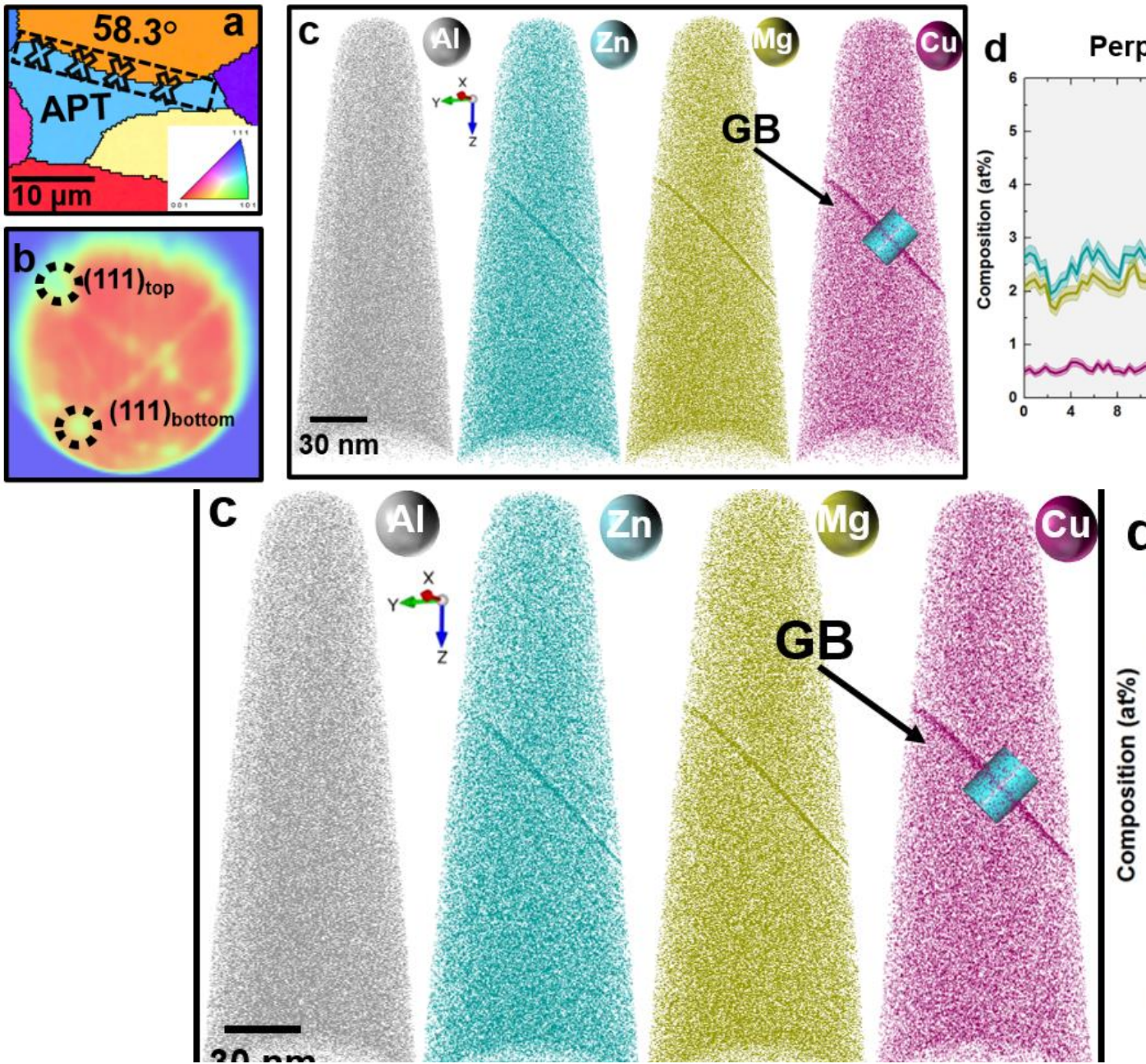

*Fig. 1. Atom probe tomography analysis for a sample in the as-quenched state: (a) EBSD IPF map showing the grain boundary (GB) for site-specific APT tip preparation; (b) Desorption map showing indexed crystallographic poles of two grains; (c) Atom maps of all elements in the as-quenched state; (d) Composition profile across the grain boundary in a 20nm-diameter cylinder. (Al, Zn, Mg, and Cu are depicted in grey, dark cyan, olive, and dark red, respectively.)*

The grain boundary is highlighted in the 3D atom map in Fig. 1(c) by the region enriched in Zn, Mg, and Cu. This observation reveals that all solute elements segregate to the grain boundary already during quenching. Quantitative measurement of the mean chemical composition across the grain boundary presented in Fig. 1(d) shows 4.6 at% Zn, 4.5 at% Mg, and 1.6 at% Cu enrichment, corresponding to a grain boundary excess (multiplier factor) of about 2 for Zn, 2 for Mg and 3 for Cu relative to the adjacent bulk solute composition. It should be noted that in the as-quenched state, the size and chemical contrast of the clusters/GP zones do not allow for a straightforward visualization by iso-composition surfaces.

Fig. 2 shows the APT analysis after aging for 0.5 hours at 120°C. Nano-sized precipitates are visible in the alloy, highlighted by 6 at% Zn iso-composition surfaces. The grain boundary is observed edge-on, shown by the precipitate enriched region as indicated by the red arrow in Fig. 2(a). Within the grain boundary, spherical precipitates enriched in Zn and Mg, are clearly revealed in Fig. 2(b). More specific, the composition profile across one grain boundary precipitate shows that it contains 60 at% Al, 20 at% Zn, 18 at% Mg and 2 at% Cu. This is close to the composition expected for GP zones ($Al_2ZnMgCu$) as previous work reported [37, 56, 59-62]. The GP zones distributed within the grain boundary are around 5 nm in size, i.e. larger than the zones with 3 nm in size formed in the adjacent bulk region.

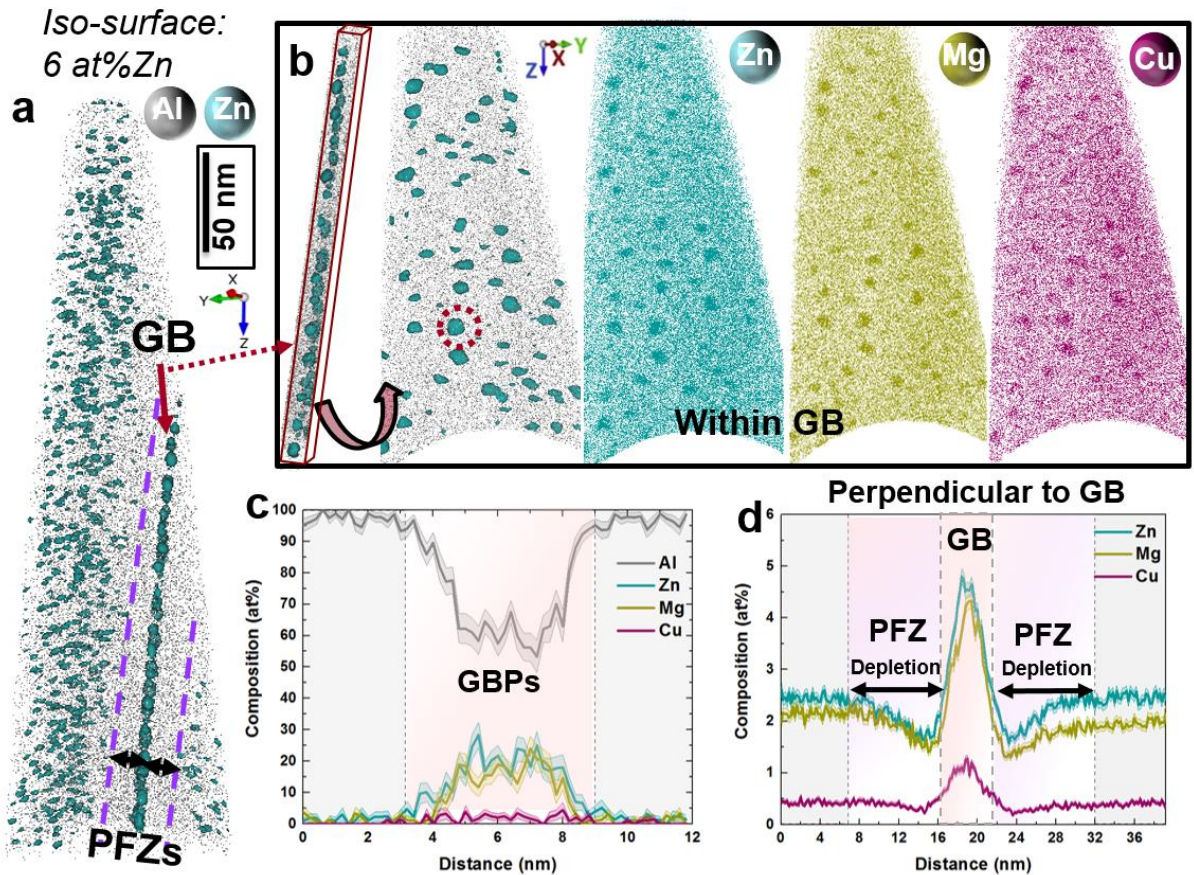

*Fig. 2. Atom probe tomography analysis after aging for 0.5 hours at 120°C. (a) Precipitates visualized by adding iso-composition surfaces of 6 at%Zn; (b) Distribution of solutes within the grain boundary (of the region shown in **a** as indicated by the red arrow); (c) 1D composition profile showing the chemistry of the selected grain boundary precipitate (GBP) (of the region shown in **b** as indicated by dashed red circle); (d) 1D composition profile across the grain boundary was computed for a region between grain boundary precipitates, along a 25nm-diameter cylindrical region-of interest. (Al, Zn, Mg, and Cu are depicted in grey, dark cyan, olive, and dark red, respectively.)*

Typical precipitate-free zones concurrently form adjacent to the grain boundary when the GP zones become visible, as indicated by the purple dashed lines in Fig. 2(a). The uniformity of the PFZs indicates a recrystallized microstructure [63]. A corresponding composition profile taken across the grain boundary in Fig. 2(d) reveals that the content of Zn and Mg is about 1.5 at% in the PFZs and is lower than inside the grain interior far away from the interface. This observation indicates that we do not only observe precipitate formation on the grain boundary but also solute depletion in the PFZs.

The precipitation behavior after aging at 120°C for 2 hours is shown in Fig. 3(a) with the grain boundary highlighted by a red rectangle. The grain boundary precipitates assume a rod-like morphology as shown in Fig. 3(b), which contrasts with the near-spherical GP zones formed in the abutting bulk. The corresponding composition with one grain boundary precipitate shows that it consists of about 61 at% Al, 21 at% Zn, 15 at% Mg and 3 at% Cu. This observation suggests that these precipitates are probably still GP zones, despite the change in morphology.

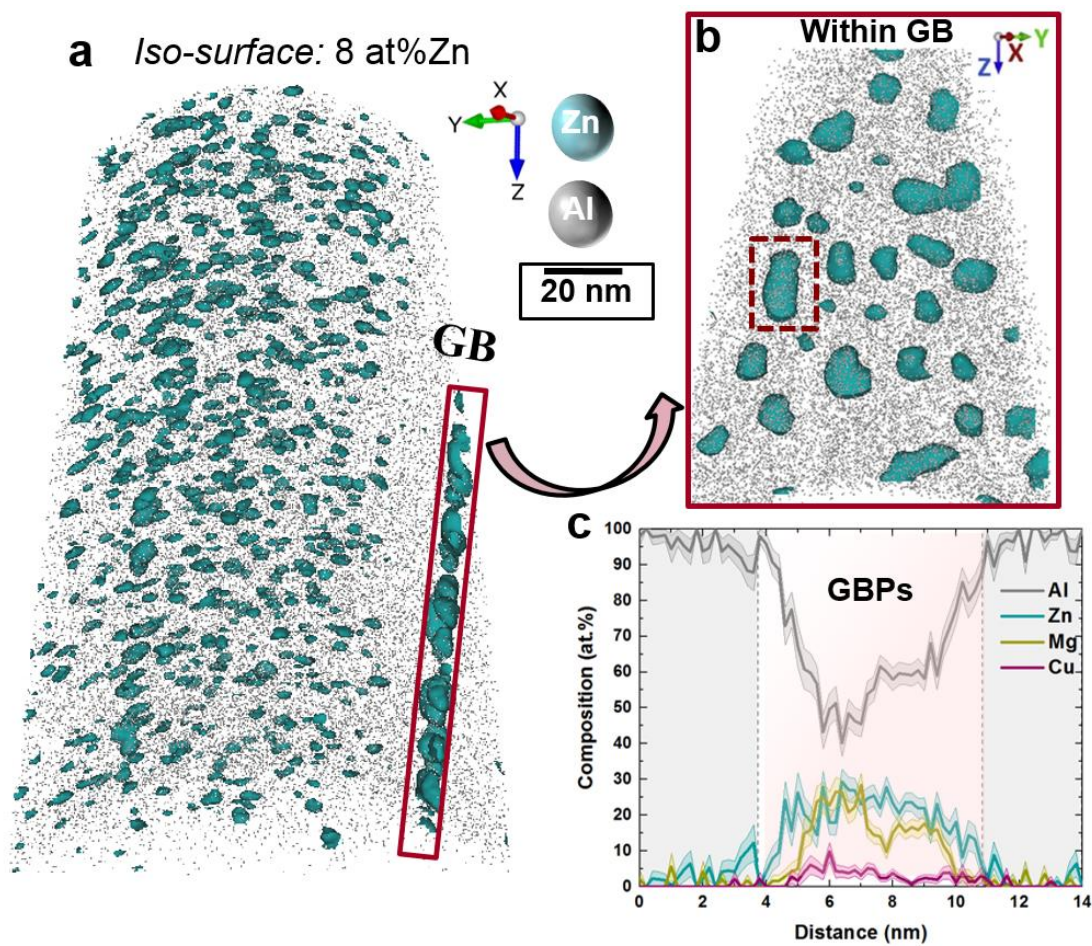

*Fig. 3. Atom probe tomography analysis showing the nanostructure after aging for 2 hours at 120°C. (a) Precipitates with 8 at% Zn iso-composition surfaces; (b) Close up map of the grain boundary precipitates (of the region shown in **a** as indicated by the solid red rectangle); (c) 1D composition profile showing the chemistry of the selected grain boundary precipitate (of the region shown in **b** as indicated by the dashed red rectangle). (Al and Zn are depicted in grey and dark cyan, respectively.)*

Fig. 4 shows the result of the APT analysis of the specimen in peak aged state, i.e. after 24 hours at 120°C. A high number density (1×10$^{24}$ m$^{-3}$) of small precipitates is homogeneously distributed in the bulk. In Fig. 4(b), a close-up on one precipitate in the bulk near the {111} planes of the Al matrix shows that its morphology is near-spherical with a diameter of around 7 nm. The corresponding composition presented in Fig. 4(c) indicates that these bulk precipitates are mainly GP zones, which profoundly contribute to strengthening.

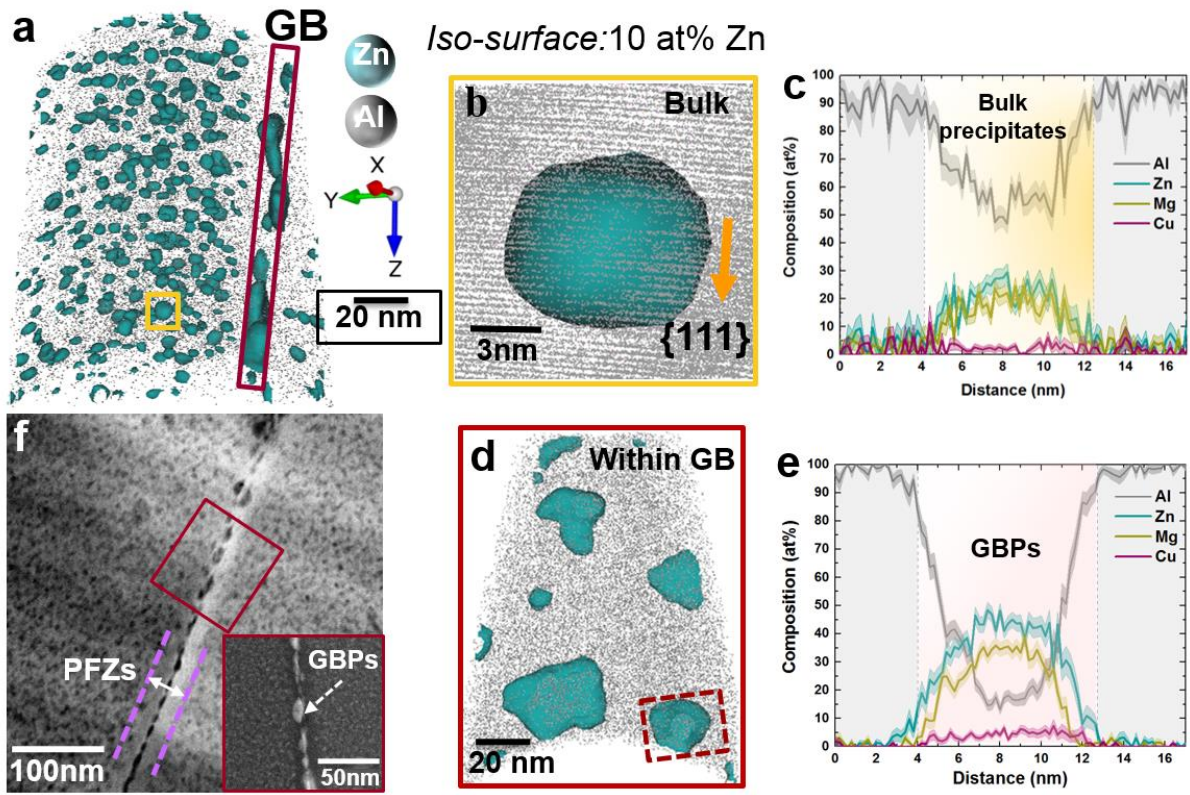

*Fig. 4. Atom probe tomography analysis taken on a specimen after aging for 24 hours at 120°C. (a) Precipitates highlighted in terms of 10 at% Zn iso-composition surfaces. (b) Close up image of one precipitate in the bulk, taken with a view along the {111} planes of the Al matrix (of the region in **a** indicated by the solid yellow rectangle); (c) 1D composition profile showing the chemistry of the selected bulk precipitate; (d) Image of the precipitates within the grain boundary (of the region shown in **a** as indicated by the solid red rectangle); (e) 1D composition profile showing the chemistry of the selected grain boundary precipitate (of the region shown in **d** as indicated by dashed red rectangle); (f) Electron micrograph shows the comparable size of the precipitates and of the PFZs in the peak aged state. (Al and Zn are depicted in grey and dark cyan, respectively.)*

Along the grain boundary, a few isolated large precipitates formed, as shown in Fig. 4(d). Most of the grain boundary precipitates grow as thick plates and are larger than 20 nm in length. The compositional evolution is accompanied by this morphology change. The chemical composition of the grain boundary precipitates shows that they contain less Al (25 at%) and more Zn (40 at%) compared to 60 at% Al and 20 at% Zn composition observed for the precipitates after 2 hours of aging at 120°C. The precipitates' composition and the plate-shaped morphology suggest that they're η' phase [37, 38, 64-66]. (As η' and η might have similar chemical compositions because the substitution of Al in the precipitates, in view of the main concern of this work, hereafter we use η' instead of making a detailed distinction between η' and η.) We assume that within the grain boundary, there is a transition of GP zones to η' plates, while in the adjacent bulk grain the GP zones prevail. The large η' precipitates nucleate and grow on the grain boundary through the accumulation of GP zones, as they have a much lower number density ($10^{15}$ m$^{-2}$) compared to the initial stage of aging ($10^{16}$ m$^{-2}$). Based on the evidence that no elongated precipitate exists within the grain boundary at this state, we suggest that the elongated precipitates observed after 2 hours of aging form as a transition phase between the small spherical GP zones and the larger η' plates.

The precipitate free zones shown in Fig. 4(a) are still noticeable in the peak aged state. In addition to the APT analysis, TEM was used to map the nanostructure in the vicinity of the grain boundary. The electron micrograph in Fig. 4(f) agrees well with the APT result in Fig. 4(a) and shows a comparable size of the coarse precipitates and the PFZs.

In the overaged state, one single large precipitate (~40 nm in length) is observed along the grain boundary as shown in Fig. 5(a). Some of the bulk precipitates show morphology changes which are characterized by a transition from the small spherical shape in the earlier stage towards coarse denser plates upon aging. The chemical composition of the coarse precipitates from the bulk show less Al (~30 at %) compared to 60 at% Al in the precipitates in the peak aged state, suggesting that they are η'. These observations indicate that in the over-aged state, some of the bulk precipitates transform from GP zones to η' precipitates. This finding is in contrast to the same phase transition that was observed for the grain boundary precipitates already in the peak aged state.

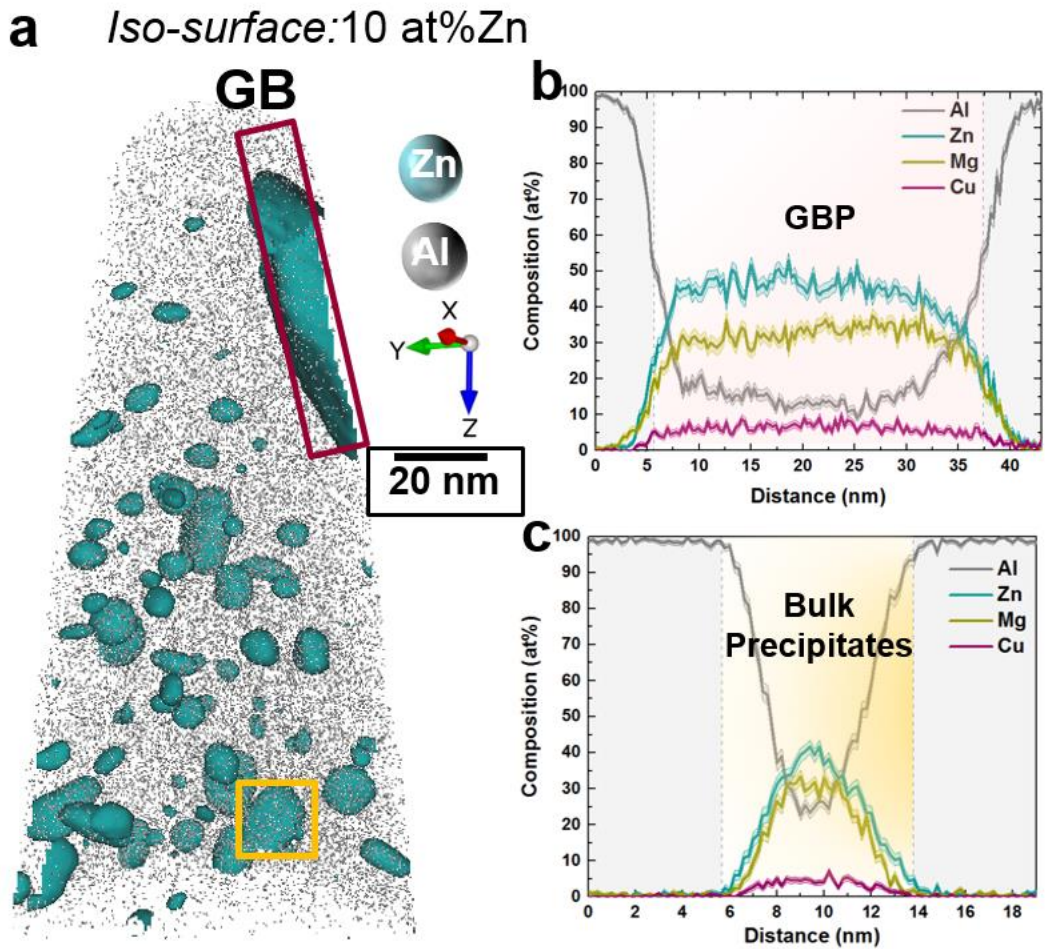

*Fig. 5. Atom probe tomography analysis in the overaged state: (a) Precipitates highlighted in terms of 10 at% Zn iso-composition surfaces; (b) 1D composition profile showing the chemistry of the selected grain boundary precipitate (of the region shown in **a** as indicated by solid red rectangle); (c) 1D composition profile showing the chemical composition of the selected bulk precipitate (of the region shown in **a** as indicated by solid yellow rectangle). (Al and Zn are depicted in grey and dark cyan, respectively.)*

# 4  Discussion

## 4.1 Quantification of the precipitation reaction

We recently described an analysis protocol for APT datasets that combines the determination of the matrix composition using nearest neighbor distributions called DIAM [67] with a radial distribution function (RDF)-based protocol to characterize the evolution of the precipitates over the course of aging. The key idea of the joint application of these two methods is that it opens up a parameter-free and versatile analysis pathway and can be applied to comparable nanostructured systems [56].

Here, this approach allows us to quantify the evolution of the precipitation sequence from three regions, namely, region I (far and unaffected from the grain boundary), region II (region adjacent to the grain boundary, essentially containing the PFZs), and region III (directly on the GB). The different regions are depicted in Fig. 6, which take the peak aged state as an example. For evaluating the precipitation state in region I, we use the information extracted form APT datasets obtained from the grain interiors discussed in more detail in ref. [56]. For regions II and III, we use cuboidal shaped regions-of-interest to split the tomographic reconstruction into two subsets, i.e., one containing the adjacent bulk after extraction of the grain boundary zone (region II), and another one containing the grain boundary only (region III). The PFZs existing in the vicinity of the grain boundary are essentially included in region II. However, it must be considered that the PFZs and the grain boundaries in the APT reconstructions may vary in size by about 10% along the GB plane so that some of the analysis features can overlap.

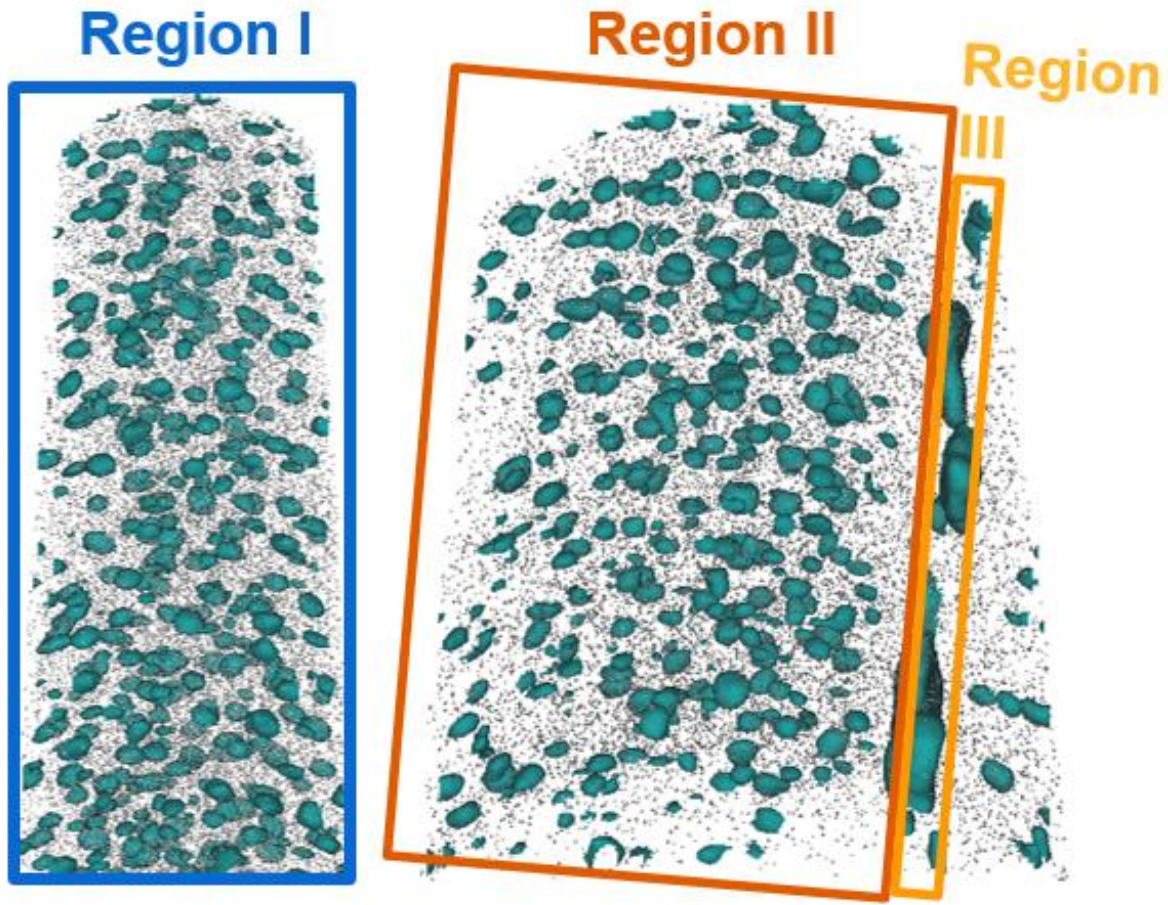

*Fig. 6. The three regions in the peak aged samples (24 hours at 120°C). The images show region I (far and unaffected from the grain boundary), region II (adjacent to the grain boundary, essentially containing the PFZs), and region III (directly on the grain boundary). (Regions I, II, and III are depicted in blue, orange, and yellow, respectively.)*

Fig. 7(a) presents the evolution of the global and matrix composition values measured in regions I and II. The global composition is derived from counting the relative number of ions of each element within the mass spectrum. The matrix composition is derived from the DIAM method [67]. In region II, the global composition of Zn eventually decreases from 2.6 to 2.3 at% in the peak aged state and more significantly to 1.7 at% in the overaged state; The global composition of Mg also show loss from 2.2 to 1.7 at% in the overaged state. This can be attributed to depletion of solutes in the PFZs enabling the segregation to be followed in region II, i.e. in diffusional reach. The observed increase in the solute composition after 0.5 hours in region II is likely related to artifacts in the investigated dataset that exhibits a large crystallographic pole. This effect is known to sometimes affect the distribution of solutes [68]. However, when excluding this specific region, we clearly see the formation of the precipitates through their consumption of solutes (global minus matrix), and the evolution of the matrix composition during aging is essentially similar in regions I and II .

Fig. 7(b) presents the evolution of the precipitates' composition during aging within regions I, II and III. In the peak aged state, the precipitates' composition in regions I and II are approx. 60 at % Al, 20 at % Zn, 18 at% Mg and 2 at% Cu, which does not reach that of the η' phase and can thus be considered as the GP zones prevail. Only when the specimen is overaged at 180 °C, the composition of the precipitates reaches a content which matches that of the η' phase. This structural transition is associated with significant capillary driven competitive particle coarsening, which also leads to a degradation of the mechanical properties [56]. It should also be noted that 2 at% Cu is comparatively richer in the very early clusters/ GP zones, indicating that Cu aids the formation of GP zones formation in the initial stages of aging as previous study also reported [41]. The Cu content (2 at%) remains unchanged after long time aging at 120°C, while it exhibits a strong increase to 5 at% in the overaged state both in regions I and II. This is explained by the low diffusivity of Cu, together with its solubility in the GP zones according to a previous study [69].

In region III, the transition from spherical GP zones to plate-shaped η' phase occur in the peak aged state as Fig. 4 shows, and the precipitates' composition presented in Fig. 7(b) also reveals less Al (45 at %), more Zn (25 at%) and Mg (25 at%) compared to the earlier stages of aging.

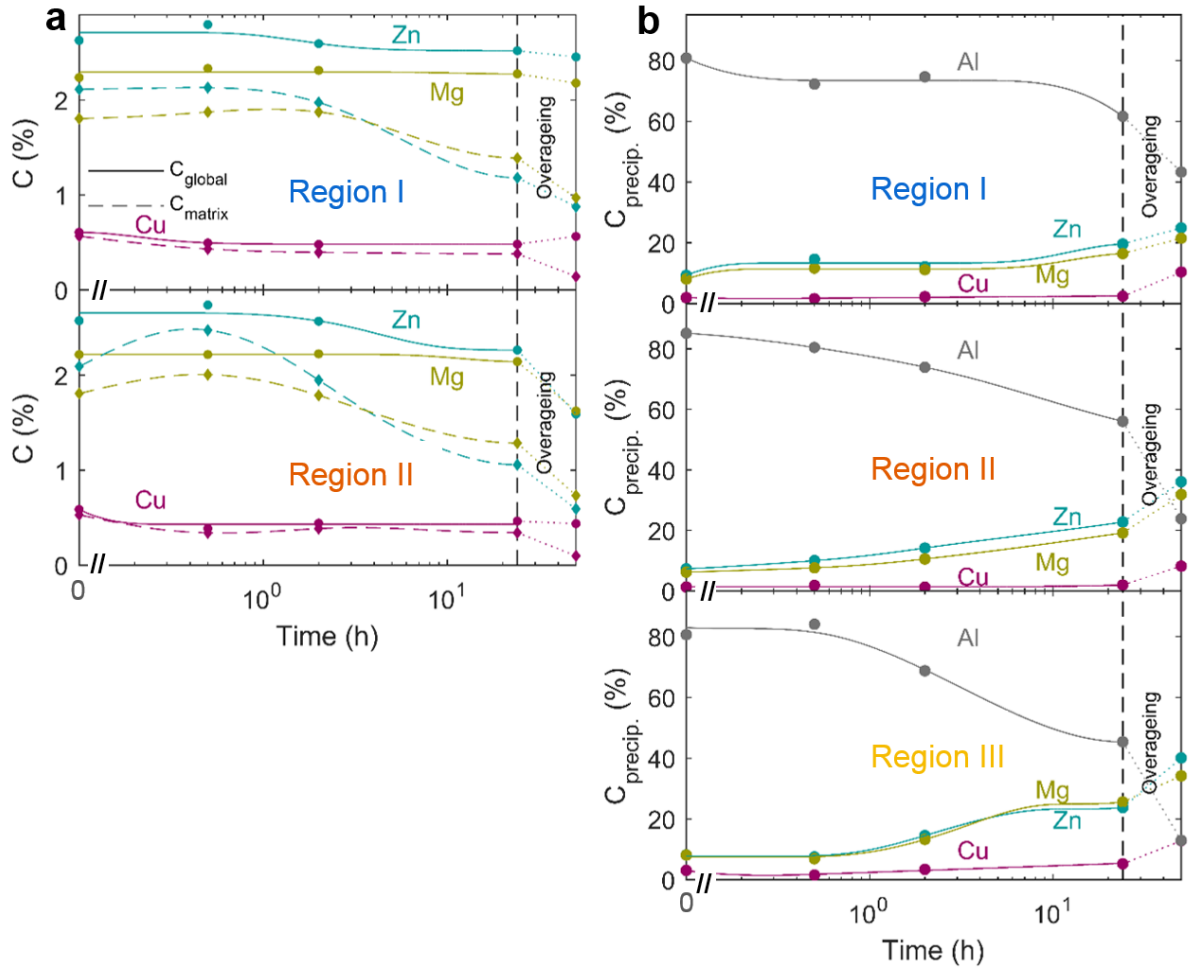

*Fig. 7. (a) Evolution of the global composition measured in the entire datasets (discs, the solid line is a guide to the eye) and the matrix composition (diamonds, the dashed line is a guide to the eye) for the different solutes as a function of aging time (b) Evolution of the precipitates' composition (discs, the solid line is a guide to the eye) for the different solutes as a function of aging time. (Regions I, II, and III are depicted in blue, orange, and yellow, respectively.)*

Fig. 8 shows the volume fraction, average radius, the number density of the precipitates, and the distance between precipitates. The results show that a large number density ($10^{25}$-$10^{26}$ m$^{-3}$) of very small (1 nm in size) solutes-rich clusters has formed already in the as-quenched state in the three regions studied. This might be caused by the natural aging during sample preparation and by the fact that real sample quenching does actually happens over a non-zero time interval which leaves some time for short relaxation diffusion.

Over the first 2 hours at 120°C aging, in regions I and II, a high number density ($10^{24}$–$10^{25}$ m$^{-3}$) of nearly spherical zones enriched in Zn and Mg form and develop into GP zones. In the same time interval, the volume fraction of precipitates increase to 1% both in the regions I and II, i.e. new precipitates formed and coarsened. A significant coarsening is shown in the overaged state as revealed by the increase of precipitate radius, volume fraction, and inter-precipitate distance. Remarkably, the precipitation kinetics, as observed in terms of the fraction, average size, and number density, are almost identical in regions I and II. This finding confirms that the region of influence that acts on the grain boundary precipitation does not extend beyond the PFZs, i.e. typically 8 – 11 nm in the current material and aging states.

We can also observe that in the initial stage of aging (0.5 hours at 120°C aging), the mean radius (~5 nm in size) of the clusters/GP zones in region III is already larger compared to those in regions I and II (~3 nm in size). During long time aging, the precipitates in region III shows a larger radius and interparticle spacing, and a lower number density.

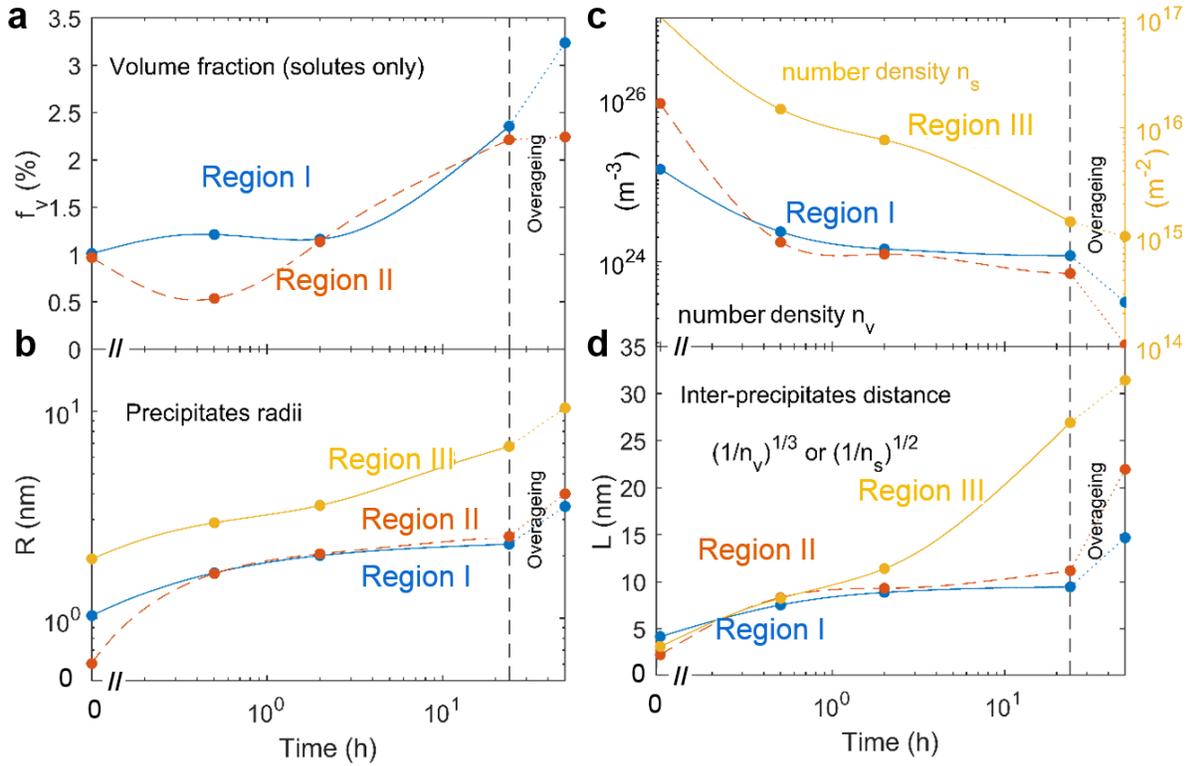

*Fig. 8. Evolution of the precipitates during aging in terms of (a) The volume fraction; (b) The radius; (c) The number density per volume in the regions I and II as shown on left axis, and precipitate number density per area in region III as shown on right axis; (d) The average distance between precipitates. (Region I, II, and III are depicted in blue, orange, and yellow, respectively.)*

## 4.2 Chemical composition on grain boundaries and within PFZs during aging

Direct analysis of the chemical composition of the grain boundaries provides additional information enabling better understanding of grain boundary segregation and precipitation. The global composition of the grain boundary regions and the available solute composition as a function of aging time were quantified separately. We analyzed the global grain boundary composition by taking the whole grain boundary volumes into account including the grain boundary precipitates. We quantified the solute composition as averaged on the grain boundaries at positions between the grain boundary precipitates, along 25nm-diameter cylindrical regions-of-interest taken across the grain boundaries.

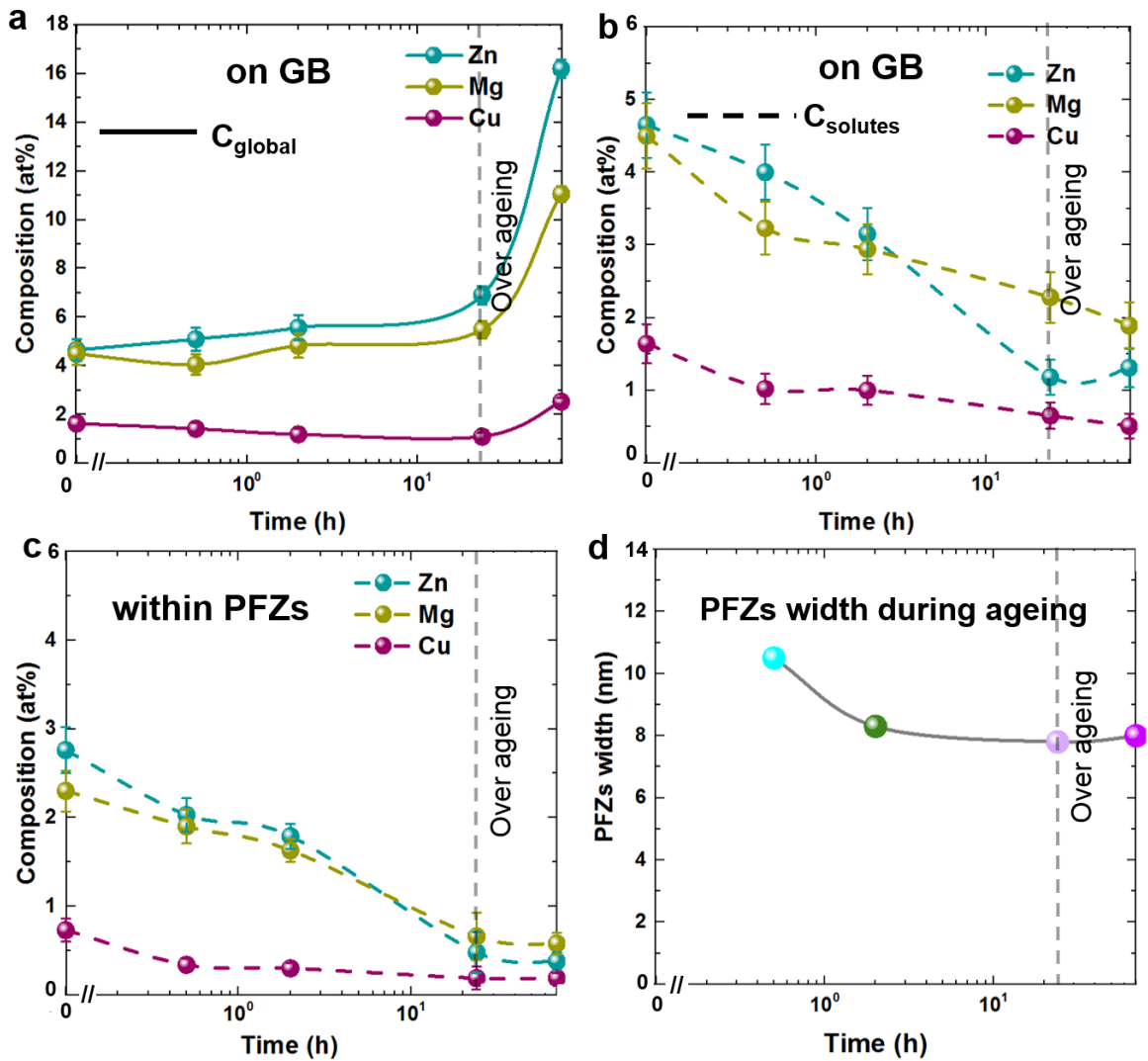

*Fig. 9. Evolution of the grain boundary composition: (a) Measured in the entire grain boundaries including the grain boundary precipitates; (b) Solute composition between grain boundary precipitates, along 25nm-diameter cylindrical regions of interest taken across the grain boundaries; (c) The available solute composition within the PFZs; (d) PFZ widths variation during aging. (Zn, Mg, and Cu are depicted in dark cyan, olive, and dark red, respectively.)*

The global composition on the grain boundaries as shown in Fig. 9(a) is characterized by strong solute segregation to grain boundary already after quenching with a grain boundary excess about 2 for Zn, 2 for Mg and 3 for Cu relative to solute composition in the bulk. The global composition increases due to more solutes from the bulk segregating to the grain boundaries during aging. The observed sharp increase in global composition in the overaged state as shown in Fig. 9(a) is most likely related to the single large precipitate imaged within the dataset collected that contained a grain boundary. This can therefore be ascribed to the small volumes sampled by APT. The solute content of the grain boundaries decreases during aging as shown in Fig. 9(b), e.g. ~5 at% Zn in the as-quenched state decreases to ~1.4 at% in the overaged state; ~5 at% Mg in the as-quenched state decreases to ~2 at% in the overaged state. This observation shows that the segregated solutes also diffuse along the grain boundaries, contributing to the growth and coarsening of the grain boundary precipitates during aging.

The PFZs appear over the entire aging process as shown in Fig. 2-5. Each individual PFZ region was extracted from the APT reconstruction using cuboidal regions-of-interest. Fig. 9(c) shows the solute composition as averaged over the probed PFZs as a function of aging time. In the as-quenched state, the solute content in the PFZs is almost at the same level as in the grain interiors. The solute composition within the PFZs show sharp drop between 2 hours and 24 hours aging at 120°C (Zn: from 1.8 to 0.5 at%; Mg: from 1.6 to 0.7 at%). The compositions of Zn and Mg within the PFZs are lower than in the grain interiors at the same aging times as shown in Fig. 7(a), e.g. by a depletion factor of 0.7 for Zn, 0.6 for Mg and 0.6 for Cu relative to the grain interior in the peak aged state. This finding indicates that the PFZs are not only devoid of any precipitates but are also solutes depleted. This result is attributed to the formation and coarsening of grain boundary precipitates which are enriched in Zn and Mg, but also to the Gibbsian segregation of solutes to the boundaries which is driven by the reduction in grain boundary energy. The insufficient diffusion range of the available matrix solute content adjacent to the PFZs and the formation of precipitates in the matrix explain this sustained solute depletion in the PFZs. The solute composition within the PFZs shows a slight decrease by a factor of ~0.2 of Zn and Mg during the over aging step. This observation suggests that only a small amount of additional solutes segregate to the grain boundaries at the later stage of aging.

The variation in the PFZs width during aging is shown in Fig. 9(d). The PFZs width are measured from the grain boundaries to the edges of the PFZs when looking edge-on at the grain boundary. The PFZs width is around 11nm at the initial state of aging (0.5 hours at 120°C), measured from the left purple line to the grain boundary as shown in Fig. 2(a). The PFZ width becomes narrower after 2 hours aging to a value of 8 nm and then remains almost constant during the later stages of aging. This result reveals that GP zones can still form in those regions of the PFZs close to the grain interiors from which solutes penetrate, driven by the composition gradient.

### 4.3 Grain boundary precipitation sequence and PFZs formation

Based on the APT results, our view of grain boundary precipitation and PFZs formation is schematically shown in Fig. 10, where solute atoms, nanoclusters, and precipitates are shown.

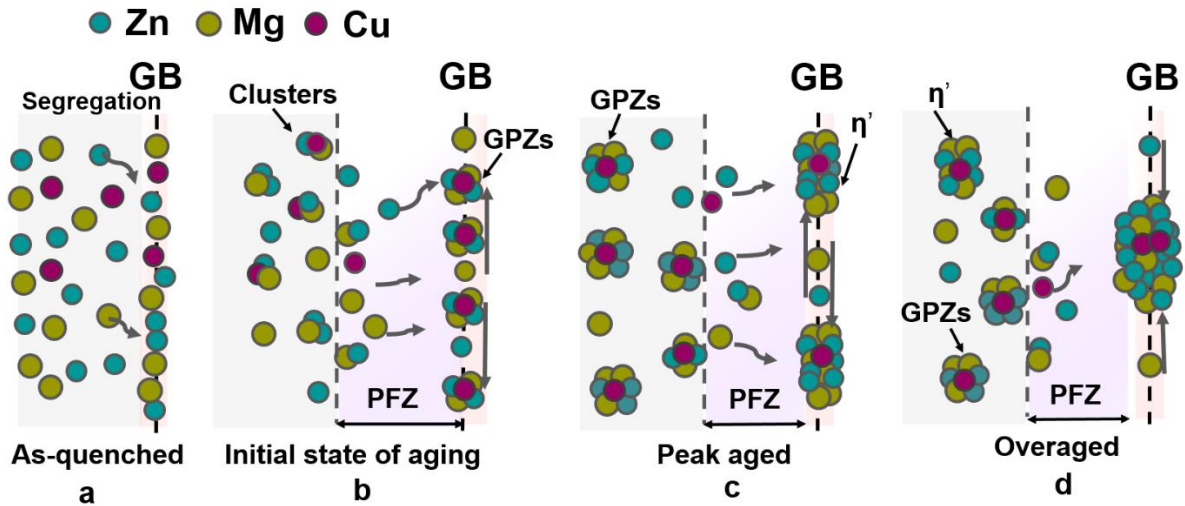

*Fig. 10. Schematic diagrams of the grain boundary precipitation and PFZ formation during aging: (a) As-quenched state; (b) Initial state of aging (0.5 hours at 120°C); (c) Peak aged; (d) Overaged. (Zn, Mg, and Cu are depicted in dark cyan, olive, and dark red, respectively.)*

In the as-quenched state, before any GP zones become visible, solutes have already segregated to the grain boundaries with a grain boundary excess of about 2 for Zn, 2 for Mg, and 3 for Cu. Solute segregation to the grain boundaries is driven by the equilibrium adsorption isotherm described originally by Gibbs and later by McLean and others [5, 70-72]. This means that the higher composition of grain boundary decorated by solute segregation is in local thermodynamic equilibrium. In the initial state of aging (0.5 hours at 120°C), the Gibbs-driven equilibrium grain boundary segregation leads to a locally higher composition of Zn and Mg, yet, at constant local chemical potential as schematically shown in Fig. 11. This local increase of solute composition brings the local composition of the grain boundaries closer to the spinodal decomposition point so that compositional fluctuation at the grain boundaries lead to earlier formation of GP zones compared to the grain interiors.

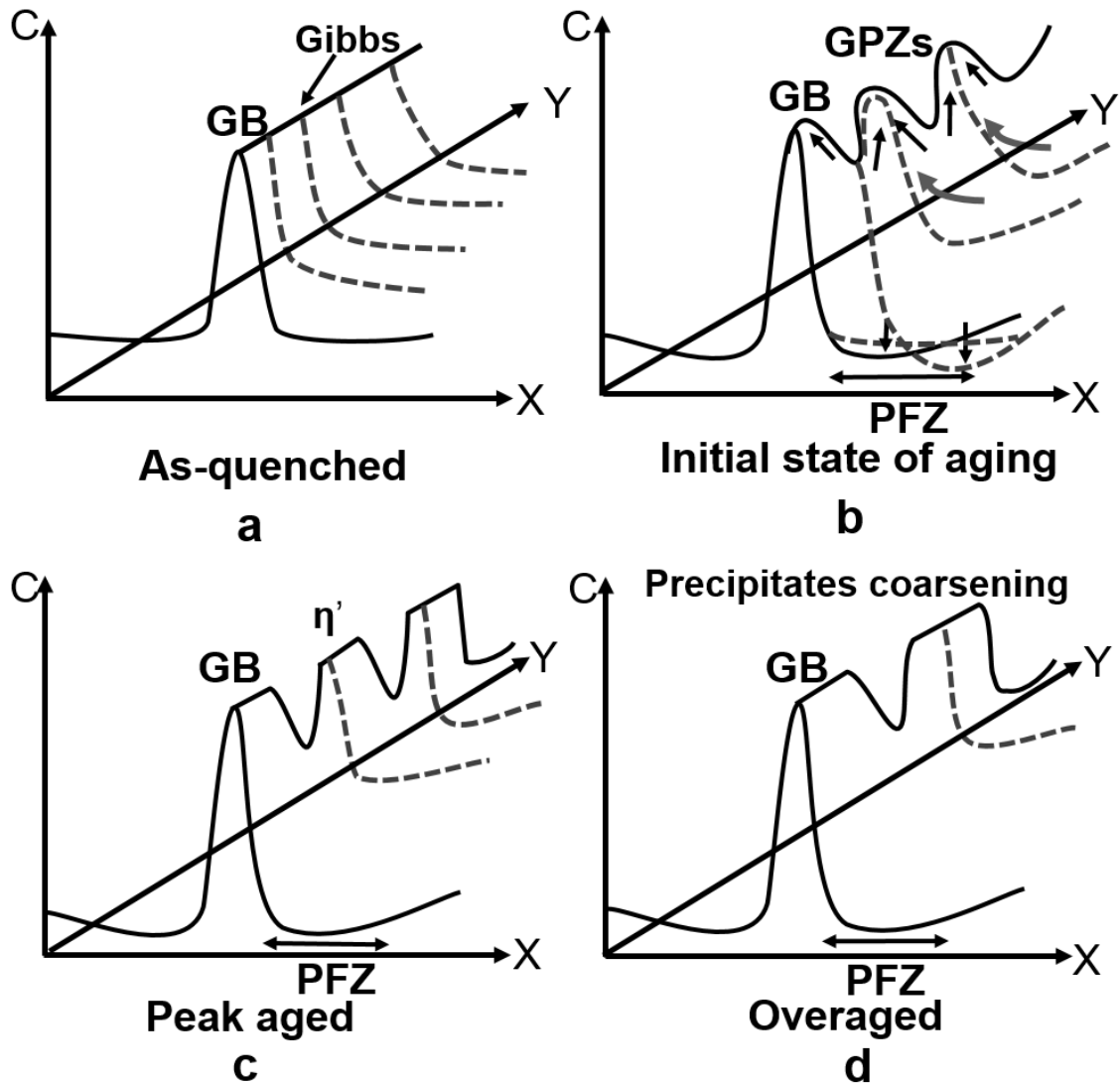

*Fig. 11. Schematic diagrams showing grain boundary composition C allowing the phase transformation during aging: (a) Segregation in the as-quenched state; (b) Phase decomposition in the initial state of aging (0.5 hours at 120°C); (c) η' formation in the peak aged state; (d) Precipitates coarsening in the overaged state.*

The GP zones distributed along the grain boundary are on average around 5 nm in size after 0.5 hours at 120°C, i.e. they are substantially larger than the GP zones with 3 nm in size formed in the bulk. The fast formation of GP zones on the grain boundary rapidly trap the solutes both from the interface itself but also from the region adjacent to the grain boundary. These processes are kinetically supported by the highly abundant quenched-in vacancies which heal out very rapidly, thereby providing additional solutes transport. This local partitioning between the Gibbs-decorated grain boundary and the region adjacent to them reduces the remaining solute composition in the near-interfacial zones. This effect kinetically reduces the availability of solutes in this region to a level insufficient to form precipitates, hence causing formation of PFZs. This means that the first step in this process is determined by a local equilibrium mechanism, i.e. driven by the Gibbs isotherm decoration of the grain boundary by Zn, Mg and Cu. The second step is also due to a local equilibrium effect, namely, the spinodal decomposition of the segregated solutes into enriched and depleted grain boundary regions. The third step, namely, the solute depletion of the regions adjacent to the interface, is related to kinetics and governed by the mass transport of those solutes that are in reach for a given time and temperature so as to follow the Gibbs adsorption isotherm and the associated confined spinodal decomposition at interfaces [73]. As mentioned above the formation of these PFZs is further influenced by the fact that the grain boundaries act as vacancy sinks, an effect which further reduces kinetics in the PFZs as suggested by Nicholson et al. [28, 31, 74-76].

With aging time increasing (between 2 hours and 24 hours at 120°C), solutes from the bulk keep segregating, forming a flux of solutes towards grain boundaries. The growth and transformation of GP zones within grain boundaries are accelerated by the solute segregation from the matrix and enhanced solute diffusion along the grain boundaries. Solute segregation creates a gradient of composition from the bulk to the grain boundaries, also promoting the growth and coarsening of grain boundary precipitates. These effects result in precipitate growth and transformation into thick η' plates with 15 nm in average length within the grain boundaries in the peak aged state, while the bulk contains mainly spherical GP zones with an average size of only 4 nm. Upon long time aging, GP zones can still form within the PFZs close to the interior of the grains with some solutes supply. This effect reduces the width of the PFZs to 8 nm compared to 11 nm in the initial state of aging.

During over aging (6 hours at 180°C after peak aged), some of the bulk precipitates transform to η' precipitates, resulting in a mixed distribution of GP zones and η' precipitates in the bulk. On the grain boundaries where the solute content exceeds that in the bulk, the growing of η' phase keeps on trapping solutes from the grain boundaries and from the matrix. This makes the η' phase grows to a size of around 40 nm in length on the grain boundaries compared to an average size of 6 nm in the bulk.

Cu has been known to be an essential addition for improving fracture toughness and stress corrosion cracking resistance in Al-Zn-Mg alloys [77-79]. In the present work, a much narrower PFZs width of ~10 nm in the Al-Zn-Mg-Cu alloy is observed compared to ~100 nm in the Al-Zn-Mg alloy in the peak aged state as previous study shown [28, 76]. We consider the reason to this is the presence of Cu, which possess attractive interactions with Zn and Mg atoms and vacancies, thus to likely enhance the formation of GP zones in the vicinity of grain boundaries and hence makes the PFZs width narrower. These soft PFZs, that can localize strain, has an adverse effect on fracture toughness and stress corrosion cracking resistance. After quenching, cold working prior to aging can also support an increase of resistance against grain boundary embrittlement since the presence of dislocation before aging might introduce a trapping of solutes in the bulk due to segregation to the dislocations [80], thus balancing the effect of segregation to the grain boundaries.

Another opportunity to improve the stress corrosion resistance and avoid the weakness of PFZs is alloying with beneficial solutes that can be rejected by the grain boundaries due to the segregation of Zn and/or Mg and/or Cu (anti-segregation). Solute segregation to the grain boundaries can influence the cohesion at grain boundaries and raise the grain boundary embrittlement susceptibility as the previous results considered [3, 7, 10, 81]. Mg segregation can also enhance hydrogen absorption as a result of the Mg-H interaction, resulting in H embrittlement and leading to the grain boundary embrittlement [10]. Anti-segregation of beneficial solutes could be an interesting strategy to occupy atomic positions within PFZs to prevent their weakening. However, more evidence is required for confirmation these suggestions.

# 5   Conclusions

Atom probe tomography has been successfully used to study the atomic-scale mechanism associated with the effects of segregation on grain boundary precipitation and PFZs formation in a model Al-Zn-Mg-Cu alloy during aging. The present observations suggest that segregation have an important influence on grain boundary precipitation. Our experimental evidence leads to the following conclusions:

1. Solute segregate to the grain boundaries with a grain boundary excess of about 2 for Zn 2 for Mg, and 3 for Cu after solution heat treated at 475°C. The local increase of solute composition brings the local composition of the grain boundaries closer to the spinodal decomposition point so that compositional fluctuation on the grain boundaries lead to earlier formation of GP zones compared to the grain interior in the initial stage of aging at 120°C .

2. With aging time increasing (between 2 hours and 24 hours at 120°C), solutes from the bulk keep segregating, forming a flux of solutes towards grain boundaries. The growth and transformation of GP zones within the grain boundaries are accelerated by the solute segregation from the matrix and enhanced solute diffusion along the grain boundaries.

3. Analysis of the ~10 nm wide precipitate-free zones (PFZs) adjacent to the enriched grain boundaries shows that the depletion zone is determined by (i) interface equilibrium segregation; (ii) formation and coarsening of the grain boundary precipitates and (iii) the diffusion range of the solutes in the matrix.

4. The precipitation kinetics, as observed by the volume fraction, size, and number density, are almost identical next to the depletion zone in the bulk and far inside the bulk grain remote from any grain boundary influence. This confirms that the region of influence of grain boundaries does not extend beyond the PFZs.

## Acknowledgements

H. Zhao would like to acknowledge the Chinese Scholarship Council for the PhD scholarship granted to support this work. U. Tezins and A. Sturm are acknowledged for their support in the use of the atom probe and PFIB facility at MPIE. The authors are grateful for the Max-Planck Society and the BMBF for the funding of the Laplace and the UGSLIT projects respectively.